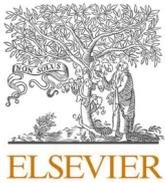
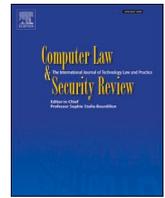

# Using sensitive data to de-bias AI systems: Article 10(5) of the EU AI act

Marvin van Bekkum

*Interdisciplinary hub for Digitalization & Society (iHub) & Institute for Computing and Information Sciences (iCIS), Radboud University, Erasmuslaan 1, 6525 GE Nijmegen NL, the Netherlands*

A B S T R A C T

In June 2024, the EU AI Act came into force. The AI Act includes obligations for the provider of an AI system. Article 10 of the AI Act includes a new obligation for providers to evaluate whether their training, validation and testing datasets meet certain quality criteria, including an appropriate examination of biases in the datasets and correction measures. With the obligation comes a new provision in Article 10(5) AI Act, allowing providers to collect sensitive data to fulfil the obligation. Article 10(5) AI Act aims to prevent discrimination. In this paper, I investigate the scope and implications of Article 10(5) AI Act. The paper primarily concerns European Union law, but may be relevant in other parts of the world, as policymakers aim to regulate biases in AI systems.

## 1. Introduction

In June 2024, the EU AI Act came into force.[1] The European Union legislator aims for, on the one hand, innovation within the European Union: introducing AI could have many competitive advantages in many areas of life, from healthcare to culture, public services, justice and climate change mitigation and adaptation.[2] The AI Act aims to 'improve the functioning of the internal market and promote the uptake of human-centric and trustworthy artificial intelligence (AI)'.[3] On the other hand, the AI Act aims to protect people against risks to health, safety, fundamental rights, and several other public interests.[4]

The AI Act includes obligations for the provider of an AI system: the 'natural or legal person that develops an AI system […] or that has an AI system […] developed and places it on the market or puts the AI system into service'.[5] Article 10 of the AI Act includes a new obligation concerning the de-biasing of AI systems: in short, the providers of AI systems must evaluate whether their training, validation and testing datasets meet certain quality criteria. Those criteria include an 'examination in view of possible biases' and 'appropriate measures to detect, prevent and mitigate possible biases.'[6] With the obligation comes a new

---

The author thanks Ylja Remmits (AlgorithmAudit), Tom Heskes (Radboud University), Frederik Zuiderveen Borgesius (Radboud University), Jaap-Henk Hoepman (Radboud University), Francien Dechesne (Leiden University) and the participants of the Leiden eLaw Conference 2024 for their useful commentary.
*E-mail address:* marvin.vanbekkum@ru.nl.

[1] Regulation (EU) 2024/1689 of the European Parliament and of the Council of 13 June 2024 laying down harmonised rules on artificial intelligence and amending Regulations (EC) No 300/2008, (EU) No 167/2013, (EU) No 168/2013, (EU) 2018/858 and (EU) 2018/1139 and (EU) 2019/2144 and Directives 2014/90/EU, (EU) 2016/797 and (EU) 2020/1828 (Artificial Intelligence Act), https://eur-lex.europa.eu/legal-content/EN/TXT/?uri=OJ:L_202401689.
[2] Recital 4 AI Act.
[3] Article 1(1) AI Act. The term 'human-centric AI' originates from earlier European Commission documents from 2019. See European Commission, *COM 2019/168 Building Trust in Human-Centric Artificial Intelligence* (2019) <https://eur-lex.europa.eu/legal-content/EN/TXT/?uri=CELEX:52019DC0168>.
[4] Recital 5 AI Act and Article 1(1) AI act.
[5] Definition of 'provider', Article 3(3) AI Act.
[6] Article 10(2)(f) and (g) AI Act.

provision in Article 10(5) AI Act, allowing providers to collect sensitive data to fulfil the obligation. The provision aims to prevent discrimination.[7] Before the AI Act, the ban on collecting sensitive data in the GDPR meant that developers were not allowed to collect such data for AI de-biasing.[8]

In this paper, I investigate the scope and implications of Article 10(5) AI Act. Practitioners implementing the AI Act, or legal scholars who wish to understand the AI Act's exception and definitions better, may benefit from the paper's analysis. The paper focuses primarily on European Union law, but may be useful in other parts of the world as well.[9] In the analysis, I will include insights from law and computer science that further elaborate the AI Act and the de-biasing of AI systems.

The contributions of the paper are as follows. A paper by Hacker discusses AI training data and Article 10 AI Act in general.[10] I focus on a different aspect of Article 10 AI Act: the exception allowing sensitive data and de-biasing. As far as I know, this is the first paper analysing the exception in-depth. The paper discusses the AI Act's definitions of AI system, provider, and deployer within the specific context of AI de-biasing. I only discuss the definitions in the AI Act as far as is necessary to understand the exception.[11] Finally, the practical difficulties providers could have to collect sensitive data fall outside of the paper's topic.[12]

The paper first discusses why providers cannot de-bias without sensitive data and an exception, in Section 2. Section 3 presents the text of the exception. Section 4 analyses the text of Article 10(5) AI Act, element by element. In Section 5, I discuss whether the aim of the exception – fighting discrimination by AI systems - is practicable for providers. Section 6 concludes.

## 2. Providers cannot de-bias without sensitive data and an exception

Before discussing the exception itself, I discuss why providers need to process sensitive data to de-bias AI systems. Next, I aim to show why providers may require an exception to the GDPR that allows AI de-biasing, based on Van Bekkum & Zuiderveen Borgesius (2023). Readers who are familiar with the topic can skip this section.[13]

### 2.1. Without sensitive data, providers cannot test proxies in data

Organisations can use AI systems to make decisions about people. For instance, a bank could use an AI system to assess whether a customer is creditworthy enough to apply for a mortgage. But using AI systems to make such decisions can lead to discrimination. For example, the bank's AI system could disproportionally deny mortgages to people with a certain ethnicity, even if the bank did not plan to discriminate.

Suppose that the developer of the bank's AI system wishes to test whether the decisions made by the AI system have effects that result in indirect discrimination of people with a certain ethnicity.[14] The developer must then first find possible biases in the AI's decisions. For such a bias test, the developer needs to know the ethnicity of the people about whom its AI system makes decisions.[15]

The developer of an AI system cannot prevent the system from having discriminatory effects against a certain ethnicity by removing explicit references to ethnicity from the AI system's design. Even if an AI system does not use explicit ethnicity data, other attributes such as a postal code can act as a *proxy*, or placeholder, for ethnicity. To find out *if* an attribute is a proxy for ethnicity with respect to an AI system's decision, the provider must collect ethnicity data. If a provider builds an AI system based on proxies that correlate strongly with ethnicity, then the user of the AI system may indirectly discriminate based on ethnicity. Roughly summarised, indirect discrimination occurs when a seemingly neutral practices ends up harming people with a certain ethnicity or other legally protected characteristic.[16]

As a fictive example, take a government organisation offering student grants to students.[17] The government organisation must check whether students live at home or not, because students who live at home receive a lower monthly grant. Students could commit fraud registering a different home address. The government organisation uses an AI system to find potential fraudsters. Based on three factors, the AI system identifies potential fraudsters: 1. type of education, 2. age and 3. distance from parent(s) house. The organisation wants to test whether the AI system, built based on these three factors, discriminates against students with a certain ethnicity.

---

[7] Recital 70 AI Act.

[8] Article 9 GDPR. See also Marvin Van Bekkum and Frederik Zuiderveen Borgesius, 'Using Sensitive Data to Prevent Discrimination by Artificial Intelligence: Does the GDPR Need a New Exception?' (2023) 48 Computer Law & Security Review 105770 <https://linkinghub.elsevier.com/retrieve/pii/S0267364922001133>.

[9] For example, the blueprint for the US AI Bill of Rights refers to 'proactive equity assessments as part of the system design'. Article 10 AI Act is an example of such a proactive assessment. See *Blueprint for an AI Bill of Rights. Making Automated Systems Work for the American People* (The White house | OSTP 2022) 5, 23, 26 <https://www.whitehouse.gov/wp-content/uploads/2022/10/Blueprint-for-an-AI-Bill-of-Rights.pdf>.

[10] Philipp Hacker, 'A Legal Framework for AI Training Data—from First Principles to the Artificial Intelligence Act' (2021) 13 Law, Innovation and Technology 257 <https://www.tandfonline.com/doi/full/10.1080/17579961.2021.1977219>.

[11] For a full paper about the definition of the term *AI system*, see e.g. Hannah Ruschemeier, 'AI as a Challenge for Legal Regulation – the Scope of Application of the Artificial Intelligence Act Proposal' (2023) 23 ERA Forum 361 <https://link.springer.com/article/10.1007/s12027-022-00725-6>. See also Theodore S Boone, 'The Challenge of Defining Artificial Intelligence in the EU AI Act' (2023) 6 Journal of Data Protection & Privacy 180 <https://ideas.repec.org/a/aza/jdpp00/y2023v6i2p180-195.html>.

[12] It is unclear if, in practice, providers will succeed in collecting the sensitive data. Organisations have several options, such as data donation campaigns, asking the data subject, or inferring from proxy attributes. For a list, see for example UK Government (Department for Science, Innovation and Technology), 'Improving Responsible Access to Demographic Data to Address Bias' (14 June 2023) <https://rtau.blog.gov.uk/2023/06/14/improving-responsible-access-to-demographic-data-to-address-bias/>.

[13] Van Bekkum and Zuiderveen Borgesius (n 8).

[14] The developer is not necessarily the user of the AI system.

[15] I Žliobaitė and B Custers, 'Using Sensitive Personal Data May Be Necessary for Avoiding Discrimination in Data-Driven Decision Models' (2016) 24 Artificial Intelligence and Law 183 <http://link.springer.com/10.1007/s10506-016-9182-5>.

[16] Frederik Zuiderveen Borgesius and others, *Non-Discrimination Law in Europe: A Primer for Non-Lawyers [PREPRINT]* (2024) <https://arxiv.org/abs/2404.08519>.

[17] The example is based on a real bias audit of the AI system of the Dutch Education Executive Agency, and the paper by Žliobaitė & Custers. See AlgorithmAudit, *Preventing Prejudice. Recommendations for Risk Profiling in the College Grant Control Process: A Quantitative and Qualitative Analysis* (Education Executive Agency of The Netherlands (DUO) 2024) <https://algorithmaudit.eu/pdf-files/algoprudence/TA_AA202401/TA_AA202401_Preventing_prejudice.pdf>. I Žliobaitė and B Custers, 'Using Sensitive Personal Data May Be Necessary for Avoiding Discrimination in Data-Driven Decision Models' (2016) 24 Artificial Intelligence and Law 183 <http://link.springer.com/10.1007/s10506-016-9182-5>.

The factors *age, education or distance* do not reveal information about the ethnicity of applicants: at first glance, the factors are neutral. To test whether the AI system discriminates against certain ethnicities, the government organisation must therefore test if the factors correlate with the ethnicities of the students.[18] Based on the collected ethnicity data, the government organisation can, for example, calculate the correlation between each of the three factors and ethnicity. Furthermore, the organisation can try to predict ethnicity using the three factors.[19] If the government organisation can use the factor (or a combination) to accurately predict ethnicity, then the factor (or combination) is a proxy, or a placeholder, for ethnicity. The government organisation can decide, based on the circumstances of the case (and perhaps fairness metrics), if the AI system discriminates against a certain ethnicity.[20]

If the government organisation had used an AI system (such as machine learning) built not upon three, but thousands of factors, the problem remains the same: the provider would still need sensitive data to decide if using such an AI system has discriminatory effects on certain ethnicities.[21]

In some cases, an audit without sensitive data may be enough. The government agency could have tested whether non-protected vulnerable groups suffer a disadvantage (for example, migrants in general).[22] But for finding discriminatory effects regarding specific ethnicities, the organisation would still need sensitive data.

### 2.2. The GDPR, in principle, prohibits providers from collecting sensitive data

In Europe, developers used to run into a problem: In principle, the GDPR prohibits the use of certain categories of personal data (also called 'sensitive data'). The GDPR's special categories of data overlap with protected characteristics in non-discrimination law. The EU non-discrimination directives prohibit discrimination based on six protected characteristics: age; disability; gender; religion or belief; racial or ethnic origin; sexual orientation. Of those six characteristics, the following four are also special categories of data in the GDPR: (1) disability, (2) religion or belief, (3) racial or ethnic origin and finally (4) sexual orientation.[23]

The GDPR itself includes one exception to the ban that is sometimes suitable for de-biasing AI systems. Organisations could ask the individual's explicit consent for collecting and using the sensitive data. But in many situations, an organisation cannot obtain valid consent of an individual. In such situations, the GDPR effectively bans using special category data for AI de-biasing.[24] Other possible exceptions in Article 9 (2) GDPR require a 'union or member state law'. As far as I know, national laws in Europe do not provide for an exception enabling the processing of sensitive data for AI de-biasing.[25]

The AI Act's exception is a European Union law within the meaning of the GDPR. The AI Act's exception is in the public interest, building on Article 9(2)(g) GDPR.[26] I use the term exception, even if dogmatically one could call the provision in Article 10 an implementation.[27] Article 9 (2)(g) GDPR allows an exception 'for reasons of substantial public interest' such as fighting discrimination, based on 'union or member state law' such as the AI Act, under three conditions. First, the law much be *proportionate to the aim pursued*. Second, the law must *respect the essence of the right to data protection*. Third, the law must *provide for suitable and specific measures to safeguard the fundamental rights and the interests of the data subject*.[28]

---

[18] In this case, the organisation could not use existing publicly available, aggregated data about ethnicity by postal code, such as the data collected by bureaus of statistics, because such information is probably not representative: the public data is not only about students who were given a grant, but about the general population. See AlgorithmAudit (n 17) 23.

[19] A factor can (individually) be non-correlated to ethnicity, but still discriminate in combination with other factors. See Aidan James McLoughney and others, '"Emerging Proxies" in Information-Rich Machine Learning: A Threat to Fairness?', *2023 IEEE International Symposium on Ethics in Engineering, Science, and Technology (ETHICS)* (IEEE 2023) <https://ieeexplore.ieee.org/document/10155045/>. Instead, predicting ethnicity is often necessary. See Keegan Hines, 'Mining for Proxies in Machine Learning Systems' (18 April 2022) <https://www.arthur.ai/blog/mining-for-proxies-in-machine-learning-systems> accessed 10 February 2025. Brian Hu Zhang, Blake Lemoine and Margaret Mitchell, 'Mitigating Unwanted Biases with Adversarial Learning' (AIES 2018) <https://dl.acm.org/doi/10.1145/3278721.3278779>. Chaudhary et al. claim to not require sensitive data by using proxies, but probably require sensitive data to validate whether the generated proxies are correct (Compare with footnote 22 of this paper). Bhushan Chaudhary and others, 'Practical Bias Mitigation through Proxy Sensitive Attribute Label Generation (PREPRINT)' (arXiv, 26 December 2023) <http://arxiv.org/abs/2312.15994>.

[20] In their assessment, the organisation could also weigh so-called *fairness metrics*, (roughly speaking) numbers which express how biased an AI system is toward different groups. Currently, it is controversial how heavy providers must weigh a fairness metric in assessing legal non-discrimination risks. See Solon Barocas, Moritz Hardt and Arvind Narayanan, *Fairness and Machine Learning* (MIT Press 2023) <https://fairmlbook.org/>.

[21] Perhaps the government organisation could try to automate or support some of the assessment with an algorithm, such as in discrimination-aware data mining (note: building such an algorithm is a challenge in itself). However, in that case, the organisation still needs sensitive data to make the algorithm and verify its output.

[22] In the real-life case this example is based on, the auditing organization tested for biases against non-protected categories, due to lack of sensitive data. Such a test could give an (initial) impression of whether the system is biased. See AlgorithmAudit (n 17). Some automated tools offer a 'bias scan' attempting to find (new) vulnerable groups in data, see Selma Muhammad, 'Auditing Algorithmic Fairness with Unsupervised Bias Discovery' (14 September 2021) <https://amsterdamintelligence.com/posts/bias-discovery> accessed 12 February 2025. Joanna Misztal-Radecka and Bipin Indurkhya, 'Bias-Aware Hierarchical Clustering for Detecting the Discriminated Groups of Users in Recommendation Systems' (2021) 58 Information Processing & Management 102519 <https://www.sciencedirect.com/science/article/pii/S0306457321000285>.

[23] Article 9(1) GDPR. See also Van Bekkum and Zuiderveen Borgesius (n 8) 5–6.

[24] M. van Bekkum & F.J. Zuiderveen Borgesius, 'Using sensitive data to prevent discrimination by artificial intelligence: does the GDPR need a new exception?', Computer Law and Security Review, Volume 48, April 2023.

[25] M. van Bekkum & F.J. Zuiderveen Borgesius, 'Using sensitive data to prevent discrimination by artificial intelligence: does the GDPR need a new exception?', Computer Law and Security Review, Volume 48, April 2023.

[26] According to Recital 70 of the AI Act, the AI Act's exception is 'a matter of substantial public interest within the meaning of Article 9(2)(g) of [the GDPR]'.

[27] In the strict legal sense, the provision in Article 10(5) AI Act implements an exception in the GDPR. However, Article 9(2)(g) GDPR would not have a legal effect for providers without the provision in Article 10(5) AI Act, so I call it an exception in this paper.

[28] I discuss whether some of these conditions hold in Section 5 of the paper.

## 3. Relevant text of the exception

**Find below, for ease of reading, the text of Article 10(5) AI Act.**

Article 10(5)

5. To the extent that it is strictly necessary for the purpose of ensuring bias detection and correction in relation to the high-risk AI systems in accordance with paragraph (2), points (f) and (g) of this Article, the providers of such systems may exceptionally process special categories of personal data, subject to appropriate safeguards for the fundamental rights and freedoms of natural persons. In addition to the provisions set out in [the GDPR] and [the GDPR for EU institutions] and [the law enforcement directive], all the following conditions must be met in order for such processing to occur:

(a) the bias detection and correction cannot be effectively fulfilled by processing other data, including synthetic or anonymised data;

(b) the special categories of personal data are subject to technical limitations on the re-use of the personal data, and state-of-the-art security and privacy-preserving measures, including pseudonymisation;

(c) the special categories of personal data are subject to measures to ensure that the personal data processed are secured, protected, subject to suitable safeguards, including strict controls and documentation of the access, to avoid misuse and ensure that only authorised persons have access to those personal data with appropriate confidentiality obligations;

(d) the special categories of personal data are not to be transmitted, transferred or otherwise accessed by other parties;

(e) the special categories of personal data are deleted once the bias has been corrected or the personal data has reached the end of its retention period, whichever comes first;

(f) the records of processing activities pursuant to [the GDPR] and [the GDPR for EU institutions] and [the law enforcement directive] include the reasons why the processing of special categories of personal data was strictly necessary to detect and correct biases, and why that objective could not be achieved by processing other data.

## 4. Analysis of the exception

The introduction of paragraph 5 of Article 10 AI Act introduces several terms that narrow the scope of the exception. I discuss these terms in the following sections: *high-risk AI system* (4.1), *provider* (4.2), *bias detection and correction and the de-biasing obligation* (4.3) and *strictly necessary* (4.4). The introduction states that the processing of the sensitive data must be *subject to appropriate safeguards for the fundamental rights and freedoms of individuals* (4.5). The exception applies 'in addition to the provisions set out in the GDPR' (4.6).

### 4.1. Exception only for high-risk AI

The exception applies to high-risk AI systems. The definition of the term *AI system* in Article 3(1) AI Act is:

> a machine-based system that is designed to operate with varying levels of autonomy and that may exhibit adaptiveness after deployment, and that, for explicit or implicit objectives, infers, from the input it receives, how to generate outputs such as predictions, content, recommendations, or decisions that can influence physical or virtual environments

For a given piece of software to qualify as an *AI system*, it must have four characteristics: autonomy, adaptiveness, an objective, and the capability to infer.[29] Any software could, in principle, show these four characteristics. The EU legislator added the characteristic 'infer' later in the legislative process, to keep the definition technologically neutral. At the very least, the definition includes machine learning and logic- and knowledge-based approaches.[30]

According to Recital 12 AI Act, the EU legislator intended to exclude 'simpler traditional software systems or programming approaches', such as Microsoft Excel's auto-sum function, from the definition. On the other hand, the definition of the term *AI system is* open-ended, where context determines if a simple system qualifies as AI or not.[31] The most decisive of the four characteristics to distinguish between 'simple software' and AI is the ability to infer, which means grossly speaking, software that obtains output from input or that derives models from input.[32] As a result, in principle, traditional rule-based (programmed) systems seem excluded from the definition, but depending on the context, the simpler systems qualify. The burden of assessing which systems are AI rests on the provider.[33] Legal certainty for providers would increase if the AI

---

[29] See Article 3 AI Act. In Recital 12 AI Act, the legislator further describes these characteristics.

[30] In a previous version of the AI Act, the Commission included a list with explicit AI techniques. In the final version, the EU legislator has removed the list and replaced it with the term 'infer', in order to keep the definition technologically neutral. The explicit list included machine learning and/or logic- and knowledge-based approaches. See David Fernández-Llorca and others, 'An Interdisciplinary Account of the Terminological Choices by EU Policymakers Ahead of the Final Agreement on the AI Act: AI System, General Purpose AI System, Foundation Model, and Generative AI' [2024] Artificial Intelligence and Law <https://link.springer.com/10.1007/s10506-024-09412-y>.

[31] See also the explanatory memorandum by the OECD, 'Explanatory Memorandum on the Updated OECD Definition of an AI System', vol 8 (2023) OECD Artificial Intelligence Papers s 2. <https://www.oecd-ilibrary.org/science-and-technology/explanatory-memorandum-on-the-updated-oecd-definition-of-an-ai-system_623da898-en>.

[32] See Recital 12 AI Act.

[33] Especially governments use mainly rule-based AI. See for more details about implications Lena Enqvist, 'Rule-Based versus AI-Driven Benefits Allocation: GDPR and AIA Legal Implications and Challenges for Automation in Public Social Security Administration' [2024] Information & Communications Technology Law 1 <https://www.tandfonline.com/doi/full/10.1080/13600834.2024.2349835>.

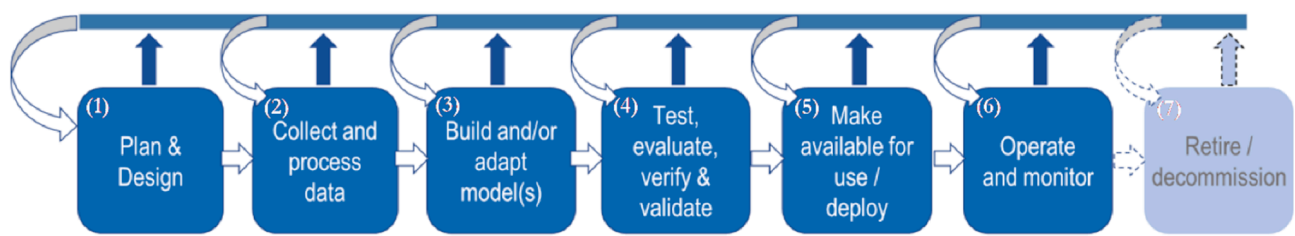

**Fig. 1.** An AI System's lifecycle according to the OECD definition. The lifecycle has seven (cyclical) stage, numbered from 1 to 7 in the figure.

---

Article 10

2. Training, validation and testing data sets shall be subject to data governance and management practices appropriate for the intended purpose of the high-risk AI system. Those practices shall concern in particular: […]

(f) examination in view of possible biases that are likely to affect the health and safety of persons, have a negative impact on fundamental rights or lead to discrimination prohibited under Union law, especially where data outputs influence inputs for future operations;

(g) appropriate measures to detect, prevent and mitigate possible biases identified according to point (f); […]

---

Office or AI Board give examples of specific use cases where simpler software qualifies as AI.[34]

The exception in Article 10(5) AI Act only applies for AI systems that the AI Act considers *high-risk*. According to Article 6 AI Act, in two cases, the AI Act considers an AI system high-risk.[35] First, the AI Act considers systems that fall in the scope of certain existing EU regulations high-risk. Second, the AI Act includes a list of high-risk AI systems in Annex III. Grossly speaking, the term high-risk encompasses AI systems in: (i) biometrics, (ii) critical infrastructures, (iii) education, (iv) employment, (v) essential private and public services, (vi) law enforcement, (vii) migration, (viii) administration of justice and democracy. The categories mention specific applications of such systems, spanning from job recruitment in the employment sector to AI systems helping judicial authorities decide on legal matters.[36] Some AI systems escape the scope of Annex III, such as dating apps.[37] There are three limited exceptions to *high-risk* in the AI Act, but the exceptions only apply if the AI system does not *profile* natural persons, which will often be the case.[38]

In sum, the definition of the term *AI system* is complex, but at the very least covers AI systems that are not too simple. Many AI systems could be high-risk in practice, especially because they profile natural persons and fall under the definition in Annex III AI Act.

*4.2. Exception applies to provider*

The AI Act's exception applies to *providers* of AI systems. According to Article 3 AI Act, a provider is 'a natural or legal person, public authority, agency or other body that develops an AI system or a general purpose AI model or that has an AI system or a general purpose AI model developed and places them on the market or puts the system into service under its own name or trademark, whether for payment or free of charge.' Several examples fall outside of the scope of the definition.[39] The provider (who places the system on the market) and developer do not have to be the same entity.[40] The exception does not apply to the deployer, the user of the AI system.[41]

What does it mean to *develop* an AI system? The OECD's definition distinguishes between different stages in an AI system's lifecycle.[42]

See the diagram depicted in Fig. 1.[43] As the diagram shows, an AI system's development consists of several stages (from the left): (1) planning and designing, (2) collecting and processing data, (3) building and/or adapting AI models, and (4) testing, evaluating, verifying and validating the AI model. (5) using/deploying and (6) operating and monitoring. It is possible to match the AI lifecycle with the definitions of provider and deployer: the provider is mostly responsible for the first four steps, while the deployer is responsible for steps 5 and 6. Where the provider *develops* the AI system and places the AI system on the market or into service, the deployer *uses* the AI system.[44]

---

[34] See Articles 64 (1) and 66(e), (g) and (i) AI Act. Article 10 AI Act comes into force on 2 August 2026, Article 113 AI Act. See also Philipp Hacker, 'Comments on the Final Trilogue Version of the AI Act [Preprint]' [2024] SSRN Electronic Journal s II <https://www.ssrn.com/abstract=4757603>. Fernández-Llorca and others (n 30). Marko Grobelnik, Karine Perset and Stuart Russell, 'What Is AI? Can You Make a Clear Distinction between AI and Non-AI Systems?' (March 2024) <https://oecd.ai/en/wonk/definition> accessed 4 February 2025.

[35] Note: the AI Act's high-risk categories do not guarantee that the system is also high-risk in other laws, such as the GDPR.

[36] Annex III AI Act.

[37] In The Netherlands, a Dutch dating app developer lacks the exception necessary to de-bias its AI system, even after being required to de-bias the AI by national authorities. See (in Dutch): College voor de Rechten van de Mens [The Netherlands Institute for Human Rights], 'Dating-app Breeze mag (en moet) algoritme aanpassen om discriminatie te voorkomen [Dating app Breeze may (and must) adjust algorithm to prevent discrimination]' (6 September 2023) <https://www.mensenrechten.nl/actueel/nieuws/2023/09/06/dating-app-breeze-mag-en-moet-algoritme-aanpassen-om-discriminatie-te-voorkomen> accessed 4 February 2025.

[38] See Article 6 AI Act. Because of the broad scope of the term *profiling*, practitioners may want to first assess whether the AI system profiles natural persons, before looking at the exceptions.

[39] The definition seems to cover all legal personhoods. An organisation renting an AI system from an AI developer is not a provider, unless the organisation also develops the AI system further. A student developing an AI system for personal use, without a name or trademark falls outside of the scope of 'provider'. This sounds logical from a product safety perspective: the moment an AI system is put on the market, its rules apply.

[40] The AI Act includes an obligation for provider and supplier to co-operate. See Article 25(4) AI Act.

[41] The deployer is 'a natural or legal person, public authority, agency or other body using an AI system under its authority except where the AI system is used in the course of a personal non-professional activity', Article 3(4) AI Act.

[42] The definition of the term *AI system* in the AI act is based on the definition created by the international organisation OECD. See Organisation for Economic Co-operation and Development (OECD), *Recommendation of the Council on Artificial Intelligence* (2023) <https://legalinstruments.oecd.org/en/instruments/oecd-legal-0449>.

[43] Organisation for Economic Co-operation and Development (OECD), *Meeting of the Council at Ministerial Level, 2-3 May 2024. Report on the Implementation of the OECD Recommendation on Artificial Intelligence, C/MIN(2024)17* (2024) fig 3.2 <https://one.oecd.org/document/C/MIN(2024)17/en/pdf>.

The terms provider and deployer are not mutually exclusive: a deployer can (also) become a provider, and a provider can (also) become a deployer. For high-risk AI systems in particular, the EU legislator intended a flexible definition of deployer. Article 25(1) AI Act states that if any deployer or third party makes a *substantial modification* or *uses the AI system for another purpose*, the AI Act considers them a provider. In such a case, the original provider becomes a third-party supplier that must cooperate with the new provider.[45] The AI Act does not define what a substantial modification or modified purpose is.[46]

*4.3. Bias detection and correction and the de-biasing obligation*

According to Article 10(5) AI Act, providers may only use the exception for 'ensuring bias detection and correction […] in accordance with paragraph (2), points (f) and (g)'. Let me cite the relevant part of Article 10 AI Act:[47]

The AI Act's exception speaks of 'detecting and correcting' biases. Comparing these terms to points (f) and (g), detecting biases means an 'examination in view of possible biases that are likely to affect the health and safety of persons, have a negative impact on fundamental rights or lead to discrimination prohibited under Union law, especially where data outputs influence inputs for future operations'. Correcting biases means taking 'appropriate measures to detect, prevent and mitigate possible biases identified according to point (f)'. Both definitions seem fairly open-ended.[48]

The focus is on *possible biases* in points (f) and (g). Neither the AI Act nor its recitals define what biases are. Dictionaries give many possible definitions for the term *bias*, such as for example: (i) 'the action of supporting or opposing a particular person or thing in an unfair way, because of allowing personal opinions to influence your judgment.' (ii) 'the fact of preferring a particular subject or thing.' (iii) 'an unfair personal opinion that influences your judgment.'[49] In scholarship, the term bias does not have a fixed definition, although many taxonomies exist.[50]

Developers may come across many types of biases during the development and use of an AI system. I name a few possible scenarios (which are not necessarily bias in the language of AI developers):

- An error made in a calculation, which leads to different results than intended. As an example: the results of the German election in Saxony had to be corrected because of a calculation error. After a correction of the error, the party AfD lost the elections in Saxony.[51]
- The results of the AI system are biased. For example, men may be overrepresented compared to women in the datasets used to train the AI system.[52]
- The AI system itself seems neutral, but the userbase is not. For example, a dating app may have less users with specific ethnicities, which results in a (possibly reinforced) underrepresentation of those users in the app.[53]

Typically, an AI developer will refer to 'bias' as a prejudice with respect to certain groups or individuals (the second example above).[54] The AI Act does not define what bias means: currently, all three of the above examples could be 'bias' in the sense of Article 10(2) AI Act.[55] In the context of this paper, which focuses on Article 10(5) AI Act, the second example (overrepresentation) makes the most sense.[56]

It follows from the introduction of Article 10(2) AI Act that detecting and correcting biases in datasets is not only possible for the provider, it is *mandatory*.[57] In Article 10(5) AI Act, the EU legislator explicitly refers to the bias detection and correction *in accordance with* Article 10(2)(f) and (g). Consequently, providers may only process sensitive data as far as strictly necessary to examine their *training, validation and testing datasets* for the bias test required by Article 10(2) AI Act. Currently, because there is no exact guidance on how to correct biases, the obligation could be broad or narrow: further guidance seems necessary.[58] The provider may not collect the data to look for biases beyond the datasets used for development, such as biases encoded in the AI model or biases that originate from applying an AI system across a different context than the one for which it was originally designed.[59]

In sum, the meaning of the term 'bias' in the AI Act seems open to interpretation, but the exception focuses on biases in the datasets used to develop the AI system, a further limitation.

*4.4. Strictly necessary*

The AI Act's exception states that providers may only process personal data for as long as 'strictly necessary'. The CJEU uses the phrase 'strict necessity' in case law about the right to the protection of personal data. The case law by the CJEU is relevant for interpreting *strict necessity* in Article 10 AI Act: Article 10 AI Act implements Article 9 GDPR, and the GDPR implements the right to the protection of personal data.[60]

---

[44] Article 3(4) AI Act.
[45] Article 25 AI Act and Recital 60 AI Act. Article 25 AI Act does not provide for possible power imbalances between provider and supplier. See also Roberta Montinaro, 'Responsible Data Sharing for AI: A Test Bench for EU Data Law' (2024) 1 EJLPT s 4 <https://universitypress.unisob.na.it/ojs/index.php/ejplt/article/view/1979>.
[46] Section 5 of the paper discusses what the lack of a clear distinction between provider and deployer means for the de-biasing obligation.
[47] Lightly edited by author.
[48] See the example in Section 2.1.
[49] Cambridge dictionary, definition of ' bias' , https://dictionary.cambridge.org/dictionary/english/bias. (accessed 4 February 2025).
[50] See, for example, S Barocas and A Selbst, 'Big Data's Disparate Impact' (2016) 104 Calif Law Rev <https://www.jstor.org/stable/24758720>. For a taxonomy, see Tilburg Institute for Law, Technology, and Society, *Handbook on Non-Discriminating Algorithms. Summary Research Report* (2021) <https://www.tilburguniversity.edu/about/schools/law/departments/tilt/research/handbook>.
[51] See (in German) Correctiv, 'Softwarefehler bei der Landtagswahl in Sachsen: Was hinter der Korrektur der Sitzverteilung steckt [Software error in the state election in Saxony: What is behind the correction of the seat distribution]' (*correctiv.org*, 6 September 2024) <https://correctiv.org/faktencheck/2024/09/06/softwarefehler-bei-der-landtagswahl-in-sachsen-was-hinter-der-korrektur-der-sitzverteilung-steckt/> accessed 4 February 2025.
[52] For example, in 2018, Amazon scrapped an AI system that was apparently biased 'because Amazon's computer models were trained […] by observing patterns in resumes submitted to the company over a 10-year period. Most came from men, a reflection of male dominance across the tech industry'. See Reuters, 'Amazon Scraps Secret AI Recruiting Tool That Showed Bias against Women' (11 October 2018) <https://www.reuters.com/article/us-amazon-com-jobs-automation-insight-idUSKCN1MK08G> accessed 4 February 2025.
[53] I discuss this example more in Section 5 (Discussion).
[54] Solon Barocas, Moritz Hardt and Arvind Narayanan, *Fairness and Machine Learning* (MIT Press 2023) <https://fairmlbook.org/>.
[55] Article 10(2)(f) AI Act refers not only to bias that possibly lead to discrimination, but also bias that could affect the 'health and safety of persons, negative impact on fundamental rights or lead to discrimination', which could be much broader.
[56] Article 10(5) AI Act aims to prevent discrimination. See recital 70 AI Act.
[57] See also, for a paper focusing on the obligation itself, Hacker (n 10).
[58] See also Section 5 (Discussion).
[59] Biases may also come from the context of use of the AI system. For possible sources of bias in AI systems, see S Barocas and AD Selbst, 'Big Data's Disparate Impact' (2016) 104 California Law Review 671.

For a given measure to be a *strict necessity*, first, the measure must be limited to what is *necessary*. Necessity means that the measure is the least intrusive measure amongst the suitable measures for the achievement of the objective it pursues. In the *Schecke* case, the CJEU ruled that a measure was unnecessary because it was 'possible to envisage measures which affect less adversely the fundamental rights of natural persons and which still contribute effectively to the objectives [...] in question [...].'[61]

The CJEU sets a high threshold for the necessity test: the measure must be limited to what is *strictly* necessary in light of its goals. The word strict means that the measure must be carefully drafted and its goals must be well-defined. According to CJEU case law regarding data retention, this requires a high level of nuance and granularity. To illustrate, in the case *Digital Rights Ireland*, the CJEU ruled that the Data Retention Directive should have been '*precisely circumscribed* by provisions to ensure that it is actually limited to what is strictly necessary.'[62] The strict necessity test is granular: the more precise and targeted a measure is, the higher the chance that the measure becomes the least restrictive option.[63]

The strict necessity test has three possible implications for providers applying the exception in Article 10(5) AI Act. First, providers will have to be precise in defining how they intend to use the sensitive data to remove biases from the dataset. This may be tricky for the purpose of AI de-biasing. While the AI Act states in Recital 70 that the exception aims to prevent discrimination, the AI Act itself does not clarify the link between bias and non-discrimination.[64] Preventing discrimination is not an exact science, unlike measuring bias. Methods for preventing discrimination are still not mature at the time of writing.[65] Providers may therefore have a tough time with the strict necessity test: it sets a high bar for a goal such as AI de-biasing. Providers will have to think carefully about which biases they aim to examine their datasets for, how this can prevent discrimination and how to carry out the examination in the least intrusive way. The state of the art in AI de-biasing, guidance by supervisory authorities and future standards to conduct the examination are also relevant in that assessment.[66]

Second, if suitable measures exist to remove the biases without using sensitive data, then those measures may be less intrusive. And it is preferable if a provider can process less data for removing the biases, for example by only using a sub-sample of the full dataset for testing and deleting the original dataset. Some techniques for measuring bias require less sensitive data than others.[67]

Finally, the provider must take into account the risk of unlawful access to the data. The longer the provider keeps the data for de-biasing, the bigger the risk of a data breach.

Overall, the strict necessity test seems to set a high bar for AI de-biasing. The future will tell how the test will be applied in AI de-biasing exactly.

### 4.5. Appropriate safeguards

While processing the sensitive data, providers must implement 'appropriate safeguards for the fundamental rights and freedoms of natural persons', which is an open norm.[68] Apart from this open norm, Article 10(5) AI Act includes a list of several conditions under Article 10 (5)(a) – (f) AI Act. For a discussion of Article 10(5)(a) AI Act, see Section 4.6.

The conditions in Article 10(5)(b)-(f) AI Act include security, privacy and accountability safeguards that the provider must implement. *Condition (b)* requires technical measures on the re-use of the data, and state-of-the-art security and privacy-preserving measures. *Condition (c)* requires confidentiality obligations with strict access controls. *Condition (d)* forbids that the data are '[...] accessed by other parties'. Presumably, the access controls and confidentiality organisations in condition (c) and other safeguards must guarantee that other parties cannot access the sensitive data. *Condition (e)* concerns data retention. *Condition (f)* concerns accountability in light of the GDPR: providers must keep a record of their processing activities, explain why processing the sensitive data is strictly necessary, and why de-biasing could not be achieved by processing other data. All of the conditions in Article 10(5)(b)-(f) overlap with the GDPR's data protection principles.[69]

The list of safeguards shows how the legislator seeks to balance data protection and non-discrimination law: the provider may process the data, but only under strict safeguards.[70] While the safeguards make sense, providers could consider more safeguards to make the exception *the least intrusive* and therefore *strictly necessary*. A trusted third party that already stores the data could perhaps share the data with provider, which means that the provider does not need to collect the data directly from data subjects. Condition (d) states that the sensitive data are not to be accessed *by* other parties: The exception does not allow providers to create, for example, entire data sharing platforms. However, it seems that the provider may still receive (anonymized) sensitive data *from* other parties that legitimately collected it, such as, for example, the national statistics bureaus.[71]

### 4.6. The GDPR still applies

According to Article 10(5) AI Act, the exception applies 'in addition to the provisions set out in [the GDPR].' We can also see this in one of the specific conditions in the exception: Article 10(5)(a) AI Act states that the exception does not apply if the provider can also fulfil the bias examination and detection effectively by processing other data, such as synthetic or anonymised data.[72] Anonymised data is non-personal, so the GDPR does not apply in that case. As the word *synthetic* suggests, synthetic data is essentially 'fake', or artificially generated data. In the context of AI de-biasing, synthetic data aims to strike a balance between privacy and usability: a dataset that is on the one hand untraceable to individuals, and on the other hand representative of the real world, so that it is still useful for detecting and correcting biases.[73] Data scientists can create representative synthetic data (either manually or by using AI) from a smaller sample of real sensitive data.[74]

---

[60] Recital 1 GDPR.
[61] *Volker und Markus Schecke GbR (C-92/09) and Hartmut Eifert (C-93/09) v Land Hessen* [2010] CJEU C-92/09 and C-93/09 [86].
[62] Emphasis added. *Digital Rights Ireland Ltd* [2014] HvJEU C-293/12 en C-594 [65].
[63] Lorenzo Dalla Corte, 'On Proportionality in the Data Protection Jurisprudence of the CJEU' (2022) 12 International Data Privacy Law 259, 270 <https://academic.oup.com/idpl/article/12/4/259/6647961>.
[64] Non-discrimination law also does not yet play well with algorithmic discrimination. See e.g. Raphaële Xenidis, 'Tuning EU Equality Law to Algorithmic Discrimination: Three Pathways to Resilience' (2020) 27 Maastricht Journal of European and Comparative Law 736, s 3.A <https://journals.sagepub.com/doi/10.1177/1023263X20982173>.
[65] See Hilde Weerts and others, 'Algorithmic Unfairness through the Lens of EU Non-Discrimination Law: Or Why the Law Is Not a Decision Tree', *2023 ACM Conference on Fairness, Accountability, and Transparency* (ACM 2023) <https://dl.acm.org/doi/10.1145/3593013.3594044>.
[66] See also Section 4.3 and for further discussion Section 5 (Discussion).
[67] For example, methods using proxy variables, unsupervised learning models, causal fairness, or synthetic data. See also section 4.6 of the paper. See also Sebastiaan Berendsen and Emma Beauxis-Aussalet, 'Fairness versus Privacy: Sensitive Data Is Needed for Bias Detection' (*UCDS research group at VU*, 14 March 2024) <https://ucds.cs.vu.nl/fairness-versus-privacy-sensitive-data-is-needed-for-bias-detection/> accessed 4 February 2025.
[68] The open norm can be filled in with the GDPR, see Section 4.6.1.
[69] See Section 4.6.1 for a comparison.
[70] See Section 2.
[71] See Sebastiaan Berendsen and Emma Beauxis-Aussalet (n 67).

Depending on whether synthetic data can be linked to an individual, synthetic data can be anonymous, and the GDPR does not apply.[75] In writing the AI Act, the EU legislator made a few 'slips of the pen' regarding synthetic data: Article 59(1)(b) AI Act explicitly names 'anonymised, synthetic or *other* non-personal data'. Article 10(5)(a) AI Act states that the exception is unnecessary if 'synthetic or anonymised data' is effective. Because these provisions equate synthetic and anonymous data, it seems as if the AI Act grants synthetic data the same status as anonymous data in law.[76] However, apart from these breadcrumbs, the AI Act is silent on the matter: further exploration of the topic seems necessary, but goes beyond the scope of the paper.[77]

Even if a provider uses (non-personal) synthetic data, the provider may still need the exception to *generate* synthetic data in the first place. Providers can only create representative synthetic data if they have a (smaller) representative sample of real synthetic data. The exception also seems useful for providers creating synthetic data.[78]

*4.6.1. Data protection principles*

The data protection principles in Article 5 GDPR also apply to the processing of sensitive data. The principles include lawfulness (Article 5 (a) GDPR), purpose limitation (Article 5(b) GDPR), data minimisation (Article 5(c) GDPR), accuracy (Article 5(d) GDPR), storage limitation (Article 5(e) GDPR), integrity and confidentiality (Article 5(f) GDPR) and finally accountability (Article 5(2) GDPR). Section 4.6 discusses the principle of lawfulness.

The principle of *purpose limitation* requires that personal data must be 'collected for specified, explicit and legitimate purposes and not further processed in a manner that is incompatible with those purposes'. Providers must clearly specify the exact purpose of the processing: which de-biasing techniques will they use, and why is sensitive data necessary for the technique? We can interpret the purpose limitation principle together with the term *strictly necessary*, which also requires a specific definition of the purposes of the processing and reflection on whether they are legitimate.[79]

The *data minimisation* principle implies that the provider must use as little data as possible for the de-biasing. In many cases, the provider could use a *subsample* of the full dataset for the de-biasing, rather than the data of all data subjects. Such a smaller sample of data has similar statistical properties as the full dataset: both must be representative.[80]

According to the *data accuracy* principle, personal data the provider gathers must not be incorrect. At first glance, for the purpose of detecting and correcting biases, the data accuracy principle seems inherently necessary: without accurate data, the bias detection and correction process would not be effective. The data must be representative. It is unclear what kind of accuracy is necessary for AI de-biasing with non-discrimination as the goal. Scholars have long debated whether non-discrimination attributes, such as ethnicity, must be self-reported by the data subject (sometimes called self-identification) or based on a predefined set of criteria.[81] There is something to be said for both approaches: ethnicity, for example, is an ambiguous term with different meanings across countries or even different contexts.[82] Depending on the definition, the ethnicity could be more or less accurate for the data subject or the AI de-biasing. As far as I know, best practices on this issue do not exist at the time of writing, but both approaches seem compatible with human rights.[83]

The *storage limitation* principle states that the provider must keep the sensitive data 'in a form which permits identification of data subjects for no longer than is necessary for the purposes for which the personal data are processed'.[84] When the provider has complied with the de-biasing obligation, the provider must delete the data, unless the provider has another compatible purpose. The provider must choose a retention period that is not longer than strictly necessary for the purpose of the processing.[85] Article 10(5)(e) AI Act rephrases the storage limitation principle quite literally: the special categories of personal data must be 'deleted once the bias has been corrected or the personal data has reached the end of its retention period, whichever comes first'.

The principle of *integrity and confidentiality* requires appropriate security of the personal data, in the form of technical and organisational measures. Article 10(5) AI Act under (b), (c), (d) and (e) give many examples of measures. Depending on the context in which the de-biasing takes place, more safeguards, such as a trusted third party, could be appropriate.[86]

---

[72] See also Carolyn Ashurst and Adrian Weller, 'Fairness Without Demographic Data: A Survey of Approaches', *Equity and Access in Algorithms, Mechanisms, and Optimization* (ACM 2023) <https://dl.acm.org/doi/10.1145/3617694.3623234>.

[73] See Section 4.4. See also Steven M Bellovin, Preetam K Dutta and Nathan Reitinger, 'Privacy and Synthetic Datasets' 22 53.

[74] Providers could create synthetic data with generative adversarial networks (GANs), for example. See, for a survey of methods, Alvaro Figueira and Bruno Vaz, 'Survey on Synthetic Data Generation, Evaluation Methods and GANs' (2022) 10 Mathematics <https://www.mdpi.com/2227-7390/10/15/2733>.

[75] See the definition of *personal data* in Article 4(1) GDPR.

[76] For example, at first glance, 'the intention of the EU Legislator to equalize anonymous and synthetic data [...] appears undeniable.' See (informally) Lorenzo Christofare - Legal status of Synthetic Data – 2023, https://www.linkedin.com/pulse/legal-status-synthetic-data-lorenzo-cristofaro.

[77] See e.g. Fontanillo López, César Augusto and Abdullah Elbi, *On Synthetic Data: A Brief Introduction for Data Protection Law Dummies* (KU Leuven 2022) <https://lirias.kuleuven.be/3838501&lang=en>. Michal S Gat and Orla Lynskey, 'Synthetic Data: Legal Implications of the Data-Generation Revolution' (2024) 109 Iowa Law Review 1087 and further.

[78] For recommendations to organisations working with synthetic data, see e.g. Payal Arora and others, *Recommendations on the Use of Synthetic Data to Train AI Models* (NU Centre, UNU-CPR, UNU Macau 2024) <https://unu.edu/publication/recommendations-use-synthetic-data-train-ai-models>.

[79] See also Section 4.4 on strict necessity.

[80] The statistical technique for finding the minimal sample is called 'power analysis'. For an example, see e.g. Jun Yu, Mingyao Ai and Zhiqiang Ye, 'A Review on Design Inspired Subsampling for Big Data' (2024) 65 Statistical Papers 467 <https://link.springer.com/10.1007/s00362-022-01386-w>.

[81] See e.g. Julie Ringelheim, 'Processing Data on Racial or Ethnic Origin for Antidiscrimination Policies: How to Reconcile the Promotion of Equality with the Right to Privacy?' [2007] SSRN Electronic Journal <http://www.ssrn.com/abstract=983685>. AI practitioners who wish to de-bias their AI also struggle with this issue. See M Andrus and others, '"What We Can't Measure, We Can't Understand": Challenges to Demographic Data Procurement in the Pursuit of Fairness' [2021], *The 2021 ACM Conference on Fairness, Accountability, and Transparency* (ACM 2021) s 6.1 <https://dl.acm.org/doi/10.1145/3442188.3445888>.

[82] The CJEU has stated that 'Ethnic origin cannot be determined on the basis of a single criterion but, on the contrary, is based on a whole number of factors, some objective and others subjective.' *Jyske Finance* [2017] CJEU C-668/15, ECLI:EU:C:2017:278 [19]. See also Sofia Jaime and Christoph Kern, 'Ethnic Classifications in Algorithmic Fairness: Concepts, Measures and Implications in Practice', *The 2024 ACM Conference on Fairness, Accountability, and Transparency* (ACM 2024) <https://dl.acm.org/doi/10.1145/3630106.3658902>.

[83] Ringelheim (n 81).

[84] Article 5(1)(e) GDPR.

[85] This also overlaps with the meaning of *strictly necessary*, see Section 4.4. The provider may still create synthetic or anonymized data, as mentioned in the introduction to this section.

[86] See Section 4.5.

Finally, the principle of *accountability* states that the providers must demonstrate compliance with all the other data protection principles. Read together with the phrase *strictly necessary*, the principle of accountability requires the provider to keep a record stating the exact purpose of the processing, explaining why the processing is truly necessary, and why the chosen measure is the *least intrusive*. Article 10 (5)(f) AI Act explicitly states a similar reading.[87]

The data protection principles can serve as a safety net in specific contexts where some of the AI Act's safeguards are not enough. The data protection principles are a flexible guide for the provider taking appropriate measures, on top of the list in Article 10(5) AI Act, which the context of the de-biasing may require.

*4.6.2. Lawful processing under the GDPR*

The principle of lawfulness in Article 5(a) GDPR also applies. Even if a provider (here, the data controller) complies with the requirements in Article 10(5) AI Act, the provider is not out of the woods yet. To process the data, the provider must have a lawful basis for the processing under Article 6 GDPR. As I mentioned before, gathering valid consent (sub a) may be difficult in many situations, because of the power asymmetries between data subject and data controller.[88]

A second candidate for a processing ground is Article 6(1)(c) GDPR juncto Article 10(2) AI Act. The AI Act states that providers *must* examine their datasets for biases. That is a concrete legal obligation. The legislator created Article 6(1)(c) GDPR for both the private and the public sector. An example from practice: banks regularly apply Article 6(1)(c) GDPR for fraud detection (as mandated by the law). The legal obligation must be directly present in a Union or member state law, which seems the case for Article 10 AI Act.[89]

Another ground for processing may be the public interest, Article 6(1)(e) GDPR, because de-biasing AI systems aims to protect people's health and safety, non-discrimination, fundamental rights. However, it seems the EU legislator originally intended that ground mainly for the public sector, not the private sector. It is unclear if a private organisation can fulfil a public interest in the sense of Article 6(1)(e) GDPR, because such organisations also have a private interest.[90]

Finally, providers could invoke ground (f): the legitimate interest ground, which the GDPR legislator originally intended only for private sector controllers.[91] In a recent case, the CJEU repeated and clarified the test for what constitutes a legitimate interest.[92] Protecting fundamental rights (such as non-discrimination) is a legitimate interest.[93] A legitimate interest is invalid if the data subject could not reasonably have expected the processing: data subjects must be informed and given a chance to object.[94] For AI de-biasing, it may be difficult for the provider to inform the data subject about the (intended) de-biasing. And if too many data subjects object, the sensitive data may become non-representative and less useful for AI de-biasing. In sum, the most plausible ground (for both sectors) seems to be Article 6(1)(c) GDPR, the legal obligation.

## 5. Discussion: is the exception effective?

The de-biasing exception and obligation target discrimination. Biases present in the data used to develop AI systems can (ultimately) have discriminatory effects on society, once the AI system has been trained and the system takes a decision. Preventing discriminatory effects of AI systems seems possible.[95] Non-discrimination law could play an important role here, even if the link between *measuring* bias and *weighing* whether a case constitutes discrimination is currently often not clear.[96]

The exception is not perfect, however. Three limitations of the exception require further discussion: (i) only providers may process sensitive data, (ii) for biases in the training, validation and testing datasets, (iii) the AI act does not describe which biases developers must remove.

(i) The exception only applies for the provider of the AI system, and the developer may not share the sensitive data with deployers, which has consequences for the effectiveness of the exception. Non-discrimination law, is a highly contextual field.[97] A developer who wishes to prevent discriminatory effects resulting from biases in data must have information about how the deployer uses the AI system in practice. Article 10 AI Act does not explicitly require deployers to work together with providers to take away discriminatory effects the systems have: the exception excludes bias monitoring.

I distinguish between three main cases of AI development and deployment: *AI as a service*, in-house development of AI, and further development of an existing AI system.[98] First, the provider and the deployer of the AI system may be entirely separate parties: the deployer rents the AI, or in other words, uses AI *as a service* (AIaaS). Multiple deployers could use an AI system created by one provider in different regions of the world, and across contexts. For example, an employment system developed by a provider in Greece could be rented to an employer (the deployer) in The Netherlands. In such cases, without further context about the setting and target group of the system, the developer cannot take away potential discriminatory effects effectively.[99] The developer and deployers must then collaborate. It seems questionable whether a single AI system can facilitate, for example, hiring in multiple countries: the developer may have to fine-tune the AI system for each individual country, in collaboration with deployers.[100] The AI Act does not require deployers of AI systems to collaborate with providers.

---

[87] Compare with Section 4.5.
[88] See Section 2.
[89] See also Christopher Kuner and others (eds), *The EU General Data Protection Regulation (GDPR): A Commentary* (Oxford University Press 2020) s Article 6.A Rationale and Policy Underpinnings <https://oxford.universitypressscholarship.com/10.1093/oso/9780198826491.001.0001/isbn-9780198826491>.
[90] ibid.
[91] ibid.
[92] *Koninklijke Nederlandse Lawn Tennisbond v Autoriteit Persoonsgegevens (KNLT v AP)* [2024] CJEU C-621/22, ECLI:EU:C:2024:857.
[93] ibid 38–40.
[94] See Recital 47 GDPR and ibid 55.
[95] Biases may ultimately also come from the context in which the AI system is deployed, rather than the data used to train the AI. However, removing biases from datasets can be a good start.
[96] See Sandra Wachter, Brent Mittelstadt and Chris Russell, 'Why Fairness Cannot Be Automated: Bridging the Gap between EU Non-Discrimination Law and AI' (2021) 41 Computer Law & Security Review 105567 <https://linkinghub.elsevier.com/retrieve/pii/S0267364921000406>. In the context of fairness metrics, it is often unclear how measuring bias and weighing discrimination interrelate. See e.g. Hilde Weerts and others, 'Algorithmic Unfairness through the Lens of EU Non-Discrimination Law: Or Why the Law Is Not a Decision Tree', *2023 ACM Conference on Fairness, Accountability, and Transparency* (ACM 2023) <https://dl.acm.org/doi/10.1145/3593013.3594044>. The problem also occurs for data scientists who try to build discrimination-aware models. See for Discrimination-aware data mining: Bettina Berendt and Soren Preibusch, 'Exploring Discrimination: A User-Centric Evaluation of Discrimination-Aware Data Mining', *2012 IEEE 12th International Conference on Data Mining Workshops* (IEEE 2012) <http://ieeexplore.ieee.org/document/6406461/>. Fairness-aware machine learning: Ioannis Pastaltzidis and others, 'Data Augmentation for Fairness-Aware Machine Learning: Preventing Algorithmic Bias in Law Enforcement Systems', *2022 ACM Conference on Fairness, Accountability, and Transparency* (ACM 2022) <https://dl.acm.org/doi/10.1145/3531146.3534644>.
[97] Michael Veale, Max Van Kleek and Reuben Binns, 'Fairness and Accountability Design Needs for Algorithmic Support in High-Stakes Public Sector Decision-Making', *Proceedings of the 2018 CHI Conference on Human Factors in Computing Systems* (ACM 2018) 10 <https://dl.acm.org/doi/10.1145/3173574.3174014>.
[98] In the EU, some sectors seem to have more in-house development of AI than others. See Charles Hoffreumon, Chris Forman and Nicolas Van Zeebroeck, 'Make or Buy Your Artificial Intelligence? Complementarities in Technology Sourcing' (2024) 33 Journal of Economics & Management Strategy 452 <https://onlinelibrary.wiley.com/doi/10.1111/jems.12586>.

Second, the provider and deployer could be the same organisation. The provider then has the necessary information for a bias examination: the sensitive data, and the information about how the *in-house* AI system is or will be deployed.

Third, sometimes, there is a grey area where a deployer becomes a provider. The deployer can become a provider when making a substantial modification, but the AI Act leaves open what substantial means. Take, as an example, a small insurance company that uses an AI system originally developed by a large AI firm, but the insurance company fine-tunes the system to target its own user-base and datasets. It is currently unclear if the insurer becomes a provider from such fine-tuning.[101] If the deployer becomes a provider, the deployer may have to remove possible biases from the data used for fine-tuning, and apply the AI Act's exception. And perhaps the provider of the original model is a 'supplier' of a high-risk AI system, and must cooperate with the fine-tuner.[102] From the point of view of de-biasing, it makes sense for the new provider to also de-bias the AI system: changes the new provider made may have introduced discriminatory effects.[103]

(ii) A second limitation of the exception is that providers must only de-bias their training, validation and testing datasets. However, biases that have discriminatory effects may originate not only from the data used to develop the AI system, but also from choices the developer makes in creating the algorithm itself and the context in which the algorithm is deployed. I give an example. A Dutch dating app ran into a problem when correcting biases. The app used a technique to recommend people based on likes from similar people (called *collaborative filtering*). It is unclear how the developer could mitigate the bias or where the bias originates from exactly: fixing the dataset may not truly prevent discrimination, because the biases may come from the userbase itself or the algorithm. Depending on the source of the bias, an advertising campaign to attract more users from certain ethnicities could be effective. Second, the developer could change the way in which the app shows users to each other (the algorithm). Both solutions do not mitigate biases resulting from the *dataset* used to develop the algorithm: they seem to be out of the scope of the exception.[104] I add that, because dating apps are not high-risk, the exception does not apply in this example anyway.[105] Limiting the bias examination to datasets seems like a conscious choice by the EU legislator: the legislator wanted to prevent bias monitoring, preventing providers from continuously processing of sensitive data to de-bias AI through its lifecycle.[106]

(iii) A final limitation is that Article 10 AI Act does not clearly state which biases providers must address and how. A recent competition involving interdisciplinary teams attempted to implement the AI Act's obligations for real-world AI systems. There is something to be said for not requiring AI providers to check for specific biases in all datasets: on the one hand, de-biasing is highly contextual, so a general obligation makes sense. On the other hand, a too generic approach 'poses the risk of missing critical vulnerabilities; domain-specific knowledge is required to identify, measure, and tackle specific biases.'[107] The European Commission and standardization organisations will have to find a middle ground between these two views. Without further guidance, especially smaller providers may lack the knowledge necessary to effectively remove discriminatory effects from their AI systems.

Overall, Article 10 AI Act could contribute to less discriminatory AI systems. Supervisory authorities could clarify which biases developers must remove to comply with the de-biasing obligation. Or perhaps a harmonized standard could clarify the de-biasing obligation: the European Commission has already requested a standard for 'specifications for appropriate data governance and data management procedures […] with specific focus on […] procedures for detecting and addressing biases and potential for proxy discrimination […].'[108] For situations in which provider and deployer are entirely separate parties, cooperation between provider and deployer seems essential to make Article 10 AI Act more effective. I think such a cooperation could be an appropriate safeguard.

## 6. Conclusion

This paper discussed the new exception allowing providers to process sensitive data for AI de-biasing (enshrined in Article 10(5) AI Act), and highlighted some of its implications.

In two influential ways, the legislator limited the scope of Article 10 (5) AI Act: (i) by writing the exception for *bias detection and correction* and leaving out *bias monitoring*, the legislator prevents the provider from monitoring biases once the AI system is finished. The focus on *bias detection and correction* prevents sensitive data from being used continuously, without data retention, and preserves necessity. (ii) The exception only applies to providers of AI systems, rather than third parties, limiting the number of parties that can store and have access to the data.

The large list of safeguards shows how the European Parliament aimed to protect sensitive data while still enabling AI de-biasing. Many of the safeguards overlap with the GDPR's data protection principles, with stricter explicit safeguards than the principles themselves may imply.

The legislator's choices result in an exception that is not a silver bullet for practitioners. Especially the two limitations in scope restrict the exception's usefulness: the exception essentially cannot be used for bias monitoring during actual use. Especially for providers of *AI as a service*, the exception seems less useful, because deployers may not collect and use sensitive data for de-biasing. Still, for providers, these limitations do not make the exception useless: providers who wish to test their datasets during the development of the AI system can still apply the exception.

---

[99] Data scientists have made this point before. See Kornel Lewicki and others, 'Out of Context: Investigating the Bias and Fairness Concerns of "Artificial Intelligence as a Service"', *Proceedings of the 2023 CHI Conference on Human Factors in Computing Systems* (ACM 2023) s 6 <https://dl.acm.org/doi/10.1145/3544548.3581463>.
[100] For example, the meaning of 'ethnicity' varies by country. See Jaime and Kern (n 82).
[101] See Section 4.2, definition of provider.
[102] See Article 25(4) AI Act.
[103] Biases can occur both upstream and downstream in the development of AI. See for an example Ryan Steed and others, 'Upstream Mitigation Is Not All You Need: Testing the Bias Transfer Hypothesis in Pre-Trained Language Models', *Proceedings of the 60th Annual Meeting of the Association for Computational Linguistics (Volume 1: Long Papers)* (Association for Computational Linguistics 2022) <https://aclanthology.org/2022.acl-long.247>.
[104] See for a discussion of this example Tim de Jonge and Frederik Zuiderveen Borgesius, 'Mitigating Digital Discrimination in Dating Apps – The Dutch Breeze Case [PREPRINT]' <https://arxiv.org/abs/2409.15828> accessed 4 February 2025.
[105] See Section 4.1 of the paper.
[106] See Section 4.2 and 4.4.
[107] Teresa Scantamburlo and others, 'Software Systems Compliance with the AI Act: Lessons Learned from an International Challenge', *Proceedings of the 2nd International Workshop on Responsible AI Engineering* (ACM 2024) <https://dl.acm.org/doi/10.1145/3643691.3648589>.
[108] See Annex 1 of European Commission, *C(2023)3215 – Standardisation Request M/593. COMMISSION IMPLEMENTING DECISION of 22.5.2023 on a Standardisation Request to the European Committee for Standardisation and the European Committee for Electrotechnical Standardisation in Support of Union Policy on Artificial Intelligence* (2023) <https://ec.europa.eu/transparency/documents-register/detail?ref=C(2023)3215&lang=en>.

Perhaps many past scandals concerning AI could have been prevented if developers had the sensitive data necessary to carefully examine their datasets. Critics of the exception should not forget that the exception essentially balances the fundamental rights to data protection and non-discrimination. The exception is one of the first attempts of any legislator worldwide to balance these two rights in the context AI de-biasing, which should be applauded.

The AI Act is long and detailed. In the coming years, different professionals must work with the AI Act's rules: government employees, consultants, auditors, researchers, lawyers, developers…the list goes ever on. Because the AI Act is complex, many practitioners struggle to determine the implications of the law's obligations. That is no surprise, because even for legal specialists, the AI Act is a complex law. Hopefully, this paper can give some support to practitioners working under the burden of the AI Act's de-biasing obligation and exception.

**Declaration of competing interest**

The authors declare that there is no conflict of interest in the creation of this work.

**Data availability**

No data was used for the research described in the article.